\documentclass[reprint,amsmath,amssymb,aps,prb,twocolumn,float]{revtex4-2}
\usepackage{graphicx}
\usepackage{graphics}
\usepackage{dcolumn}
\usepackage{bm}
\usepackage{amsfonts}
\usepackage{amssymb}
\usepackage{float}
\usepackage{xcolor}
\usepackage{soul}
\usepackage{natbib}
\usepackage{braket}
\usepackage{hyperref}

\definecolor{green}{rgb}{0,0.6,0.1}

\newcommand{\NTU}{Department of Physics and Center for Theoretical Physics, National Taiwan University, Taipei 10617, Taiwan\looseness=-1}
\newcommand{\NCTS}{Physics Division, National Center for Theoretical Sciences, Taipei 10617, Taiwan\looseness=-1}
\newcommand{\IAMS}{Institute of Atomic and Molecular Sciences, Academia Sinica, Taipei 10617, Taiwan\looseness=-1}

\begin{document}

%Title of paper
%\title{Large shift current enhancement via sub-bandgap and charge-neutral exciton excitations in BN nanotubes and single BN layer}
\title{Large shift current via in-gap and charge-neutral exciton excitations in BN nanotubes and single BN layer}

\author{Yi-Shiuan Huang$^1$}      
%\email{www.tommy2@gmail.com} 
\author{Yang-Hao Chan$^{2,3}$}        \email{yanghao@gate.sinica.edu.tw}
\author{Guang-Yu Guo$^{1,2}$}         \email{gyguo@phys.ntu.edu.tw} 
\affiliation{$^1$\NTU} 
\affiliation{$^2$\NCTS}
\affiliation{$^3$\IAMS}

\date{\today}

\begin{abstract}
We perform {\it ab initio} many-body calculations to investigate the exciton shift current 
in small diameter zigzag BN nanotubes and also single BN sheet, 
using the GW plus Bethe-Salpeter equation (GW-BSE) method with the newly developed efficient algorithms.
Our GW-BSE calculations reveal a giant in-gap peak in the shift current spectrum
in all the studied BN systems due to the excitation of the A exciton.
The peak value of the excitonic shift current is more than three times larger than
that of the quasiparticle shift current, and is attributed to the gigantic enhancement 
of the optical dipole matrix element by the A exciton resonance. The effective exciton 
shift current conductivity is nearly ten times larger than the largest shift conductivity observed 
in ferroelectric semiconductors. Importantly, the direction of the shift current in the BN 
nanotubes is found to be independent of the tube chirality ($n,0$) (or diameter), contrary
to the simple rule of $ sgn(J_\text{shift})=\text{mod}(n,3)$ predicted by previous 
model Hamiltonian studies. Finally, our {\it ab initio} calculations also show that 
the exciton excitation energies decrease significantly with the decreasing diameter due to 
the curvature-induced orbital rehybridization in small diameter zigzag BN nanotubes. 
\end{abstract}

\maketitle

\section{INTRODUCTION}
Shift current is one of the primary mechanisms for the bulk photovoltaic effect (BPVE) 
(also known as the photogalvanic effect), which can generate DC photocurrent in 
noncentrosymmetric crystals due to their second order optical response to light irradiation. 
The shift of the real space Wannier charge center of the excited electron is responsible 
for the shift current \cite{Kraut1979,Sturman1992,Aversa1995,Sipe2000}. 
In contrast to the conventional photovoltaic effect, shift current is a bulk phenomenon 
that does not require a p-n junction to separate the 
optically generated electron-hole pairs for a DC photocurrent. 
Consequently, the BPVE can be exploited to generate the above band gap photovoltage~\cite{Bhatnagar2013}
and thus to fabricate high power-conversion efficiency solar cells~\cite{Green2016,Cai2017}.
Therefore, there has been a resurge of interest in the BPVE in recent years.

An exciton is a type of collective excitation formed by a bound electron-hole pair interacting 
via Coulomb interaction, and can be optically generated. Although an exciton itself is charge-neutral, 
the process of exciton excitation can generate a shift current called exciton shift current 
due to Wannier charge center shift of the electron and hole in noncentrosymmetric crystals~\cite{Morimoto2016}. 
Furthermore, the exciton shift current can have exotic sub-bandgap peaks owing to 
the exciton binding energy~\cite{Morimoto2016, Chan2021}. Recently, the sub-bandgap exciton shift current 
was observed in the noncentrosymmetric semiconductor CdS by using THz emission spectroscopy 
at the low temperature of 2 K\cite{Sotome2021}. The amplitude of the exciton shift current 
is comparable to the shift current driven by free electron-hole pair excitation.

BN nanotubes (BN-NTs) are formed by rolling up a single hexagonal BN sheet (see Fig. 1) 
along a specific chiral vector $(n,m)$.~\cite{Saito1998} There are three types of BN-NTs, 
namely armchair $(n,n)$ nanotubes, zigzag $(n,0)$ nanotubes (Fig. 1), 
and chiral $(n,m)$ nanotubes where $ n \neq m $ \cite{Saito1998}. 
However, there is no second-order nonlinear optical (NLO) response (e.g., second-harmonic
generation and BPVE) in armchair BN-NTs~\cite{Guo2005b}. Moreover, previous experiments~\cite{Lee2001}
indicate that among the grown BN-NTs, the zigzag structure is usually favored.
Therefore, we focus on zigzag BN-NTs in this paper.

BN-NTs exhibit large many-body interaction effects 
owing to their one-dimension (1D) and wide bandgap nature, as have been demonstrated by 
previous {\it ab initio} many-body theory studies at the level of GW 
plus Bethe-Salpeter equation (GW-BSE) \cite{Wirtz2006,Park2006}. 
Furthermore, huge diameter-dependent A exciton peaks have been observed in BN-NTs \cite{Yu2009,Li2010},
being consistent with the theoretical predictions~\cite{Wirtz2006,Park2006}.
Therefore, we expect zigzag BN-NTs to be ideal candidates for observing the large subbandgap exciton 
shift current due to their following two properties \cite{Sotome2021}. 
First, exciton absorption spectra in BN-NTs are well separated from the continuum of free electron-hole 
excitation owing to their significantly renormalized optical spectra and large exciton 
binding energy~\cite{Wirtz2006,Park2006}. 
Second, strongly bounded excitons with a large binding energy impede thermal dissociation 
into free electron-hole pairs \cite{Wirtz2006,Park2006}. 

\begin{figure}[tbph] \centering
\includegraphics[width=8.5cm]{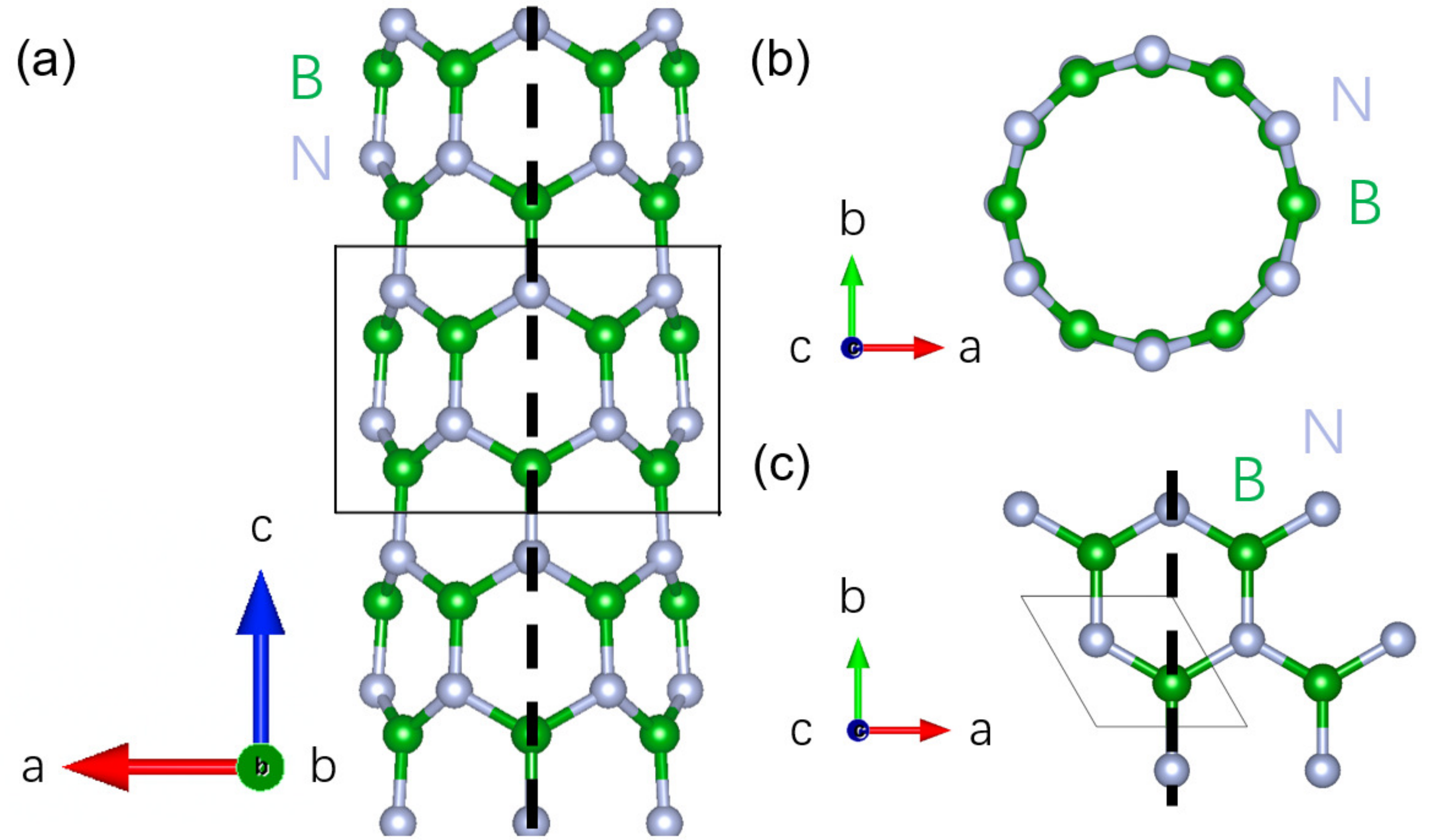}
\caption{Crystal structures of the zigzag BN-NT (6,0), and the hexagonal BN sheet. 
(a) Side and (b) top views of the (6,0) BN-NT. (c) Top view of the single BN sheet. 
The solid box in (a) indicates the unit cell. Dashed lines in (a) and (c) denote the mirror planes.}
\label{fig:struc}
\end{figure}

However, {\it ab initio} studies of the BPVE in BN-NTs
at GW-BSE level have not been reported, mainly because such {\it ab initio}
calculations are technically and computationally challenging, as will be explained below.
Nevertheless, simple model Hamiltonian calculations without~\cite{Kral2000} and with~\cite{Konabe2021} 
the inclusion of the the excitonic effect have been performed,
predicting that zigzag BN-NTs have shift current along the tube axis 
with the direction determined by the chiral index ($n$,0). 
In particular, the shift current direction would follow the simple rule 
of sgn($J_\text{sh})=\text{mod}$($n$,3) \cite{Kral2000,Konabe2021}. 
According to this rule, 
in a bundle of zigzag BN-NTs, 1/3 of them would have the shift current along the positive tube axis,
another 1/3 would possess the current flowing in the opposite direction and the other 1/3 would
have the zero current. That is to say, a bundle of zigzag BN-NTs would have either zero or very small net shift current,
i.e., zigzag BN-NTs would not be suitable for applications in photovoltaic solar cells and nonlinear optoelectronic devices. 
On the other hand, the prediction of this simple rule is theoretically surprising. 
A symmetry analysis (see Sec. II. A below) indicates that the shift current would flow along the
$y$-axis in a single BN sheet [see Fig. 1(c)], which is also the tube axis when the BN sheet is rolled 
up to form a zigzag BN-NT. Consequently, for zigzag BN-NTs with a large diameter, the shift current
would always be along the positive tube axis since the curvature effect 
would be very small~\cite{Guo2005a,Guo2005b}, i.e., the above mentioned rule would not occur. 
In this context, it is important to perform the state-of-the-art {\it ab initio} calculations
to investigate this detrimental prediction. 

In this work, therefore, we perform the state-of-the-art {\it ab initio}  GW-BSE calculations
of the shift current in small diameter zigzag BN-NTs [$(5,0),(6,0),(7,0),(8,0)$] as well as single BN layer.
Because of possible large curvature effect on the optical properties~\cite{Guo2005a,Guo2005b}, the shift current
in small BN-NTs could depend significantly on the tube diameter. On the other hand, the 
optical properties of large diameter BN-NTs would be rather similar to 
that of the single BN sheet~\cite{Guo2005a,Guo2005b}, and thus are not considered here.
Among other things, our {\it ab initio} calculations reveal that the shift current
due to the exciton excitations will be dramatically enhanced. Furthermore, the direction of the shift current
calculated without and with the excitonic effect included, is always along the $c$-axis, 
i.e., being independent on the tube index ($n,0$) (or tube diameter).
Therefore, our work demonstrates that the zigzag BN-NT bundles are 
promising materials for high power conversion efficiency solar cells as well as high sensitivity photodectors.

We would like to comment that extending the existing {\it ab initio} GW-BSE approach to calculate 
the exciton shift current is not straightforward \cite{Chan2021}. In fact, there has been only 
one reported fully {\it ab initio} study on the exciton shift current \cite{Chan2021}. 
In Ref.~\cite{Fei2020}, to account for the excitonic effect, Fei {et al.} used 
the linear optical coefficients derived from their {\it ab initio} GW-BSE calculations. 
However, since the Coulomb interaction between the electron and hole in an exciton 
is not explicitly taken into account, such calculations are still based on 
an independent-particle approximation \cite{Fei2020} and hence are not a fully {\it ab initio} GW-BSE approach. 
In Ref. \cite{Chan2021}, to take the strong excitonic effect in two-dimensional (2D) materials
into account, a time-dependent adiabatic GW (TD-aGW) approach was developed and used to calculate the exciton shift current.
Although the {\it ab initio} TD-aGW method can properly include the excitonic effect 
on the optical responses \cite{Chan2021},  the extremely high computational cost 
of conducting real-time propagation prevents it from studying complex structures such as BN-NTs. 
In this work, we thus develop a computationally efficient approach that combines the GW-BSE 
and sum-over-state formalism derived from the perturbative density-matrix approach 
within the mean-field approximation \cite{Taghizadeh2018} to calculate the exciton shift current. 

The rest of this paper is organized as follows.
In Sec. II, we introduce the crystal structure of the BN-NTs and the BN sheet, 
followed by a brief description of the theories and computational details used in this work.
In particular, the computationally efficient approach mentioed above for the exciton shift current calculations
will be outlined. The main results are presented in Sec. III.
In Sec. III A, we present the calculated electronic and optical properties
of the BN sheet. The distinguished features in the optical spectra are
analysed in terms of the electronic band structure and interband optical transition matrix elements.
In Sec. III B, the electronic properties of the BN-NTs are reported, which will be
used to understand the calculated optical absorption and shift current spectra in subsections that follows.
In Secs. III C and III D, the calculated optical absorption and shift current spectra of the BN-NTs
are presented, respectively.
Finally, the conclusions drawn from this work are summarized in Sec. IV.

\section{CRYSTAL STRUCTURES AND COMPUTATIONAL METHODS}

\subsection{Symmetry and shift conductivity tensor}
Single hexagonal BN sheet is a noncentrosymmetric crystal with space group $ P\overline{6}m2 $ 
and point group $D_{3h}$. A symmetry analysis would show that the single BN sheet has three 
nonzero shift conductivity tensor elements $ \sigma^{xxy} $, $ \sigma^{yxx} $, and $ \sigma^{yyy} $, 
where the first Cartesian index is the direction of the current and the second and third indice 
are the polarization directions of the external field. 
Furthermore, $\sigma^{xxy}=\sigma^{yxx}=-\sigma^{yyy}$ \cite{Guo2005b,Boyd2003}, 
i.e., there is only one inequivalent nonzero element. 
As an example, let us consider the difference between in-plane shift conductivity tensor 
elements $\sigma^{xxx} $ and $ \sigma^{yyy} $ in the single BN sheet [see Fig. 1(c)]. 
Figure 1(c) indicates that the atomic structures on the left and right regions to the mirror plane
(denoted by the dashed line) (the $y$-axis) are symmetric. 
Therefore, the shift current along the direction normal to the mirror plane (the $x$-axis)
would be zero if the second order combination of external fields is even under the mirror symmetry, 
i.e., $\sigma^{xxx} = 0$. On the other hand, such mirror symmetry does not exist in the direction along the
mirror plane (the $y$-axis), and thus the conductivity element $\sigma^{yyy}$ would not be zero. 

When a hexagonal BN sheet is wrapped up to form a zigzag BN-NT with chiral index ($n,0$),
the $y$-axis would become the tube axis [the $c$-axis, see Fig. 1(a)]
and the point group of the resultant BN-NT ($n$,0) is $ C_{2nc}$. 
Thus, the conductivity tensor element $\sigma^{ccc}$ is nonzero. 
On the other hand, there is no azimuthal shift current in the zigzag BN-NTs because
it corresponds to the direction of the $x$-axis in the single BN sheet (see Fig. 1).

\subsection{Density functional theory calculations}
{\it Ab initio} calculations, based the density functional theory (DFT) 
with the local density approximation (LDA), 
are performed to determine the ground state properties of the considered zigzag BN-NTs 
and also the single BN sheet. A supercell geometry is adopted to simulate a BN-NT in which
the nanotubes are arranged in a square array with a minimum distance 
of 12 \r{A} between the neighboring nanotubes. A slab supercell method is used to model the single BN sheet,
and the inter-sheet distance used is over 16 \r{A}. 
In the structural optimization calculations, the accurate projector-augmented wave (PAW) method plus 
the conjugate gradient approach, as implemented in
the VASP package~\cite{Kresse1993,Kresse1996}, is used to determine the atomic positions 
and lattice constants of BN-NTs. 
A large plane-wave cutoff energy of 450 eV is adopted. Theoretical equilibrium structures
are obtained when the forces acting on all the atoms and the uniaxial stress were less than 
0.005 eV/\AA$ $ and 2.0 kBar, respectively.

The ground electronic structure calculations are performed by using the plane-wave pseudopotential
method as implemented in the Quantum Espresso package \cite{Giannozzi2009}. 
The optimized norm-conserving Vanderbilt pseudopotential \cite{Hamann2013} is exploited here. 
The $1 \times 1 \times 32$ and $18 \times 18 \times 1$ 
Monkhorst-Pack $k$-grids \cite{Monkhorst1976} are used to evaluate the Brillouin zone (BZ) integrals 
for the BN-NTs and the single BN sheet, respectively. The energy cutoff for the plane-wave 
basis set is 50 Ry. The resultant electronic structures are used in the subsequent GW-BSE calculations,
as described below. 

\subsection{Quasiparticle band structure calculations}

The present GW-BSE calculations are performed via the BerkeleyGW 
package \cite{Hybertsen1986,Deslippe2012,Rohlfing2000}.
The quasiparticle energy bands are calculated by solving the Dyson equation,
\begin{equation}
[-\frac{1}{2}\nabla^2+V_{\mathrm{ion}}+V_{\mathrm{H}}+\Sigma(E^{\mathrm{QP}}_{n\bf{k}})]\psi^{\mathrm{QP}}_{n\bf{k}}=E^{\mathrm{QP}}_{n\bf{k}}\psi^{\mathrm{QP}}_{n\bf{k}},
\end{equation}
where $\Sigma$,  $E^{\mathrm{QP}}_{n\bf{k}} $, and $ \psi^{\mathrm{QP}}_{n\bf{k}} $ are the self-energy operator,
the energy, and the wave function of the quasiparticles within the $ G_0W_0 $ approximation, 
respectively \cite{Deslippe2012}. 

In the present one shot $G_0W_0$ calculations~\cite{Jornada2017}, a nonuniform neck subsampling (NNS) $k$-grid 
of $1 \times 1 \times 8$ ($18 \times 18 \times 1$) with a subsampling of 10 points in the mini-Brillouin zone, 
600 (1600) bands, and a dielectric cutoff energy of 50 (50) Ry for the BN-NTs (the single BN sheet) are used. 
%one shot $ G_0W_0 $ calculations \cite{Jornada2017}. 
Truncation of the Coulomb interactions between the BN-NT (the BN sheet) and its periodic images 
is implemented \cite{Ismail-Beigi2006}. 
The dynamic dielectric matrix is computed within the independent particle approximation (IPA) 
and Hybertsen-Louie generalized plasmon pole model \cite{Hybertsen1986}.

\subsection{Exciton excitation calculations}
The exciton wavefunction can be expressed as the linear combination of the quasiparticle electron-hole pairs,
\begin{equation}
\Psi_s(\textbf{r}_e,\textbf{r}_h)=\sum\limits_{\textbf{k},m,n}A^s_{mn\textbf{k}}\psi_{\textbf{k},n}(\textbf{r}_e)\psi^*_{\textbf{k},m}(\textbf{r}_h),
\label{eq2}
\end{equation}
where band index $m$ ($n$) sums over the valence (conduction) bands only.
Exciton envelope function of the $s$-th exciton state $ A^s_{mn\textbf{k}}$ can be obtained by solving the BSE,
\begin{equation}
(E^{\mathrm{QP}}_{n\textbf{k}}-E^{\mathrm{QP}}_{m\textbf{k}})A^s_{mn\textbf{k}}+\sum\limits_{m'n'\textbf{k'}}\braket{mn\textbf{k}|K^{eh}|m'n'\textbf{k'}}=\Omega^sA^s_{mn\textbf{k}}
\end{equation}
where $\Omega^s$ and  $E^{\mathrm{QP}}_{m\textbf{k}} $ ($ E^{\mathrm{QP}}_{n\textbf{k}} $) are 
the $s$-th exciton excitation energy and the valence (conduction) band quasiparticle excitation 
energies, respectively. $ K^{eh} $ is the electron-hole interaction kernel, which includes an exchange 
repulsive bare Coulomb term and a direct electron-hole attractive screened Coulomb 
term \cite{Deslippe2012,Rohlfing2000}. The imaginary part of the dielectric function including 
the excitonic effect can be expressed as
\begin{align}
\varepsilon^{"}_{\mathrm{BSE}}&=\frac{g_s\pi e^2}{\epsilon_0 N_kV_c}\sum\limits_{s}|
\sum\limits_{mn\textbf{k}}A^s_{mn\textbf{k}}\textbf{x}_{mn\textbf{k}}\cdot\textbf{e}|^2\delta(\omega-E^s)\nonumber\\
&=\frac{g_s\pi e^2}{\epsilon_0 N_kV_c}\sum\limits_{s}|\textbf{X}_s\cdot\textbf{e}|^2\delta(\omega-E^s)
\label{eq4}
\end{align}
where $ \textbf{x}_{mn\textbf{k}}\cdot\textbf{e}$ is the dipole matrix element along the polarization direction $\textbf{e}$ and $g_s$ accounts for the spin degeneracy.
$N_k$ is the number of $k$ points and $V_c$ is the unit cell volume. 
The exciton dipole matrix element $\textbf{X}_s $ is defined as
\begin{equation}
\textbf{X}_s\equiv\sum\limits_{mn\textbf{k}}A^s_{nm\textbf{k}}\textbf{x}_{mn\textbf{k}}.
\end{equation}
When neglecting the excitonic effect, optical excitations are given by direct transitions 
between quasiparticle electron-hole pairs. Within the IPA approximation, the imaginary part of the dielectric function 
reduces to \cite{Rohlfing2000},
\begin{equation}
\varepsilon^{"}_{\mathrm{IPA}}=\frac{g_s\pi e^2}{\epsilon_0 N_kV_c}\sum\limits_{mn\textbf{k}}|\textbf{x}_{mn\textbf{k}}\cdot\textbf{e}|^2\delta(\omega-(E^{\mathrm{QP}}_{n\textbf{k}}-E^{\mathrm{QP}}_{m\textbf{k}})).
\label{eq6}
\end{equation}
Equation (\ref{eq4}) has the same structure with Eq. (\ref{eq6}) except that the dipole matrix element 
is replaced by the exciton dipole matrix element.
Our BSE calculations for the BN-NTs (the single BN sheet) are computed on a dense $k$-grid of 
$1 \times 1 \times 64$ ($72 \times 72 \times 1$) with the dielectric cutoff of 10 (10) Ry. 
%Over $2n$ (4) conduction bands and $2n$ (4) valence bands are taken into account in the present BSE calculations 
%for the ($n$, 0) BN-NT (the single BN sheet).

\subsection{Exciton shift current calculations}
The shift current density along the $a$-axis is given by~\cite{Sipe2000}
\begin{equation}
J^a_{sh}(\omega)=2\sum\limits_{bc}\sigma^{abc}(0;\omega,-\omega)E^b(\omega)E^c(-\omega),
\end{equation}
where $ \sigma^{abc} $ is the third-rank conductivity tensor, $b$ and $c$ denote the polarization directions 
of the electric fields of the incident light. 
In the IPA, $\sigma^{abc}$ can be expressed as~\cite{Sipe2000,Ibanez-Azpiroz2018} 
\begin{multline}
\sigma^{abc}(0;\omega,-\omega)=-\frac{i\pi e^3}{4N_kV_c}\sum\limits_{n,m,\textbf{k}}f_{nm}\\
(\textbf{x}^b_{mn}[\textbf{x}^{c}_{nm}]_{;k_a}+\textbf{x}^c_{mn}[\textbf{x}^{b}_{nm}]_{;k_a})[\delta(\omega_{mn}-\omega)+\delta(\omega_{nm}-\omega)],
\label{eq8}
\end{multline}
where $\textbf{x}^a_{nm}$ and $[\textbf{x}^{a}_{nm}]_{;\textbf{k}}$ are the dipole matrix element and 
its generalized momentum derivative, respectively. The generalized momentum derivative of quantity $O_{mn}(\textbf{k})$ is given by
\begin{equation}
[O^b_{nm}(\textbf{k})]_{;k_a}=\partial_{ka}O^b_{nm}(\textbf{k})-i[\xi^a_{nn}(\textbf{k})-\xi^a_{mm}(\textbf{k})]O^b_{nm}(\textbf{k})
\end{equation}
where $\xi^a_{nm}(\textbf{k})$ is the $a$ component of the Berry connection.~\cite{Sipe2000}
$ f_{nm}=f_n-f_m $ and $ \hbar\omega_{nm}=E_n-E_m $ where
$f_n$ denotes the occupation factor and $E_n$ is the energy of the $n$th band at the $\textbf{k}$ point.
Note that here band indices $m$ and $n$ should sum over all the states. 

%\subsection{Exciton shift current calculations}
As mentioned before, to include the excitonic effect, we use the efficient perturbative density-matrix approach to calculate $ \sigma^{abc} $ 
[see Eq. (B1a) in Ref. \cite{Taghizadeh2018}]. In this case, $\sigma^{abc}$ can be written as

\begin{widetext}
\begin{align}
\sigma^{abc}(0,\omega,-\omega) & =-\frac{g_s e^{3}}{2V}\sum_{ss'}\left[ \frac{X_{s}^{b}\Pi_{ss'}^{a}X_{s'}^{c*}}{\left(\hbar\omega-E_{s}+i\eta\right)\left(\hbar\omega-E_{s'}-i\eta\right)}+\frac{V_{s}^{a*}X_{ss'}^{b}X_{s'}^{c}}{-E_{s}\left(\hbar\omega-E_{s'}+i\eta\right)}+\frac{V_{s}^{a}X_{ss'}^{b*}X_{s'}^{c*}}{-E_{s}\left(-\hbar\omega-E_{s'}-i\eta\right)}\right]\nonumber\\ 
 & +\left(b\leftrightarrow c,\omega\leftrightarrow-\omega\right),
\label{eq9}
\end{align}
\end{widetext}
where we define $V^a_s\equiv\sum_{cv\mathbf{k}}A^{s*}_{cv\mathbf{k}}v^a_{cv\mathbf{k}}$ with velocity matrix elements $v^a_{cv\mathbf{k}}$, and inter-exciton coupling matrix elements, $X_{ss'}$ and $\Pi_{ss'}$ are defined as,
\[
\Pi_{ss'}^{a}\equiv\sum_{cvc'\mathbf{k}}A_{cv\mathbf{k}}^{s}v_{c'c\mathbf{k}}^{a}A_{c'v\mathbf{k}}^{s'*}-\sum_{cvv'\mathbf{k}}A_{cv'\mathbf{k}}^{s}v_{v'v\mathbf{k}}^{a}A_{cv\mathbf{k}}^{s'*},
\]
and $X_{ss'}^{b}=Y_{ss'}^{b}+Q_{ss'}^{b}$. We note that $c$ and $v$ denote that the summation runs over the conduction and valence band, respectively. Here, $\textbf{X}_{ss'}$ is an ill-defined operator. However, it can be separated into the well-defined $ \textbf{Y}_{ss'} $ 
(interband part) and $ \textbf{Q}_{ss'} $ (intraband part) operators, where
\[
Q_{ss'}^{b}=i\sum_{c'v'\mathbf{k'}}A_{c'v'\mathbf{k}'}^{s*}\left(A_{c'v'\mathbf{k}'}^{s'}\right)_{;k_{b}'},
\]
and
\[
Y_{ss'}^{b}=\sum_{\substack{c'\neq c_{1}\\v'\neq v_{1}}}\sum_{c_1 v_1\mathbf{k}}A_{c'v'\mathbf{k}'}^{s*}\left[A_{c_{1}v'\mathbf{k}'}^{s'}x_{c'c_{1}}^{b}-A_{c'v_{1}\mathbf{k}'}^{s'}x_{v_{1}v'}^{b}\right].
\]

We develop a post-processing program to implement this formalism,
which is then used to calculate the excitonic shift current conductivity using the outputs from the BerkeleyGW package.

It is not straightforward to evaluate the intraband portion of the position operator 
$\textbf{Q}_{ss'} $. $ \textbf{Q}_{ss'} $ involves the numerical derivative of 
the exciton envelope function $ A^s_{nm\bf{k}} $ with respect to $ \textbf{k} $. 
Furthermore, $ A^s_{mn\bf{k}} $ has an arbitrary $k$-dependent gauge. To solve the problem, 
we use the locally smooth gauge adopted in Ref. \cite{Chan2021} to calculate the 
intraband position operator. The idea of the locally smooth gauge is to rotate the wave functions 
at neighboring  $k$ points in such a way that the overlap of connected wave functions 
is Hermitian \cite{Chan2021,Souza2004,Virk2007}.

In this work, the $\delta$-function is approximated by a Gaussian function with a 0.1 eV broadening. 
For the BN NTs, we use the effective unit cell volume rather than the supercell volume. 
The effective unit cell volume is given by $ V_c=\pi [(D/2+d/2)^2-(D/2-d/2)^2]T=\pi DdT $, 
where $D$ is the tube diameter, $T$ is the length of translational vector and $d$ is the effective thickness 
of the nanotube walls, which is set to the interlayer distance (3.28 Å) of $h$-BN \cite{Guo2005a}.
For the single BN sheet, the effective unit cell volume is $ V_c=A_cd$ where 
$A_c$ and $d$ are the area of the unit cell and effective thickness of the sheet, respectively. 

\begin{figure}[tbph] \centering
\includegraphics[width=8.0cm]{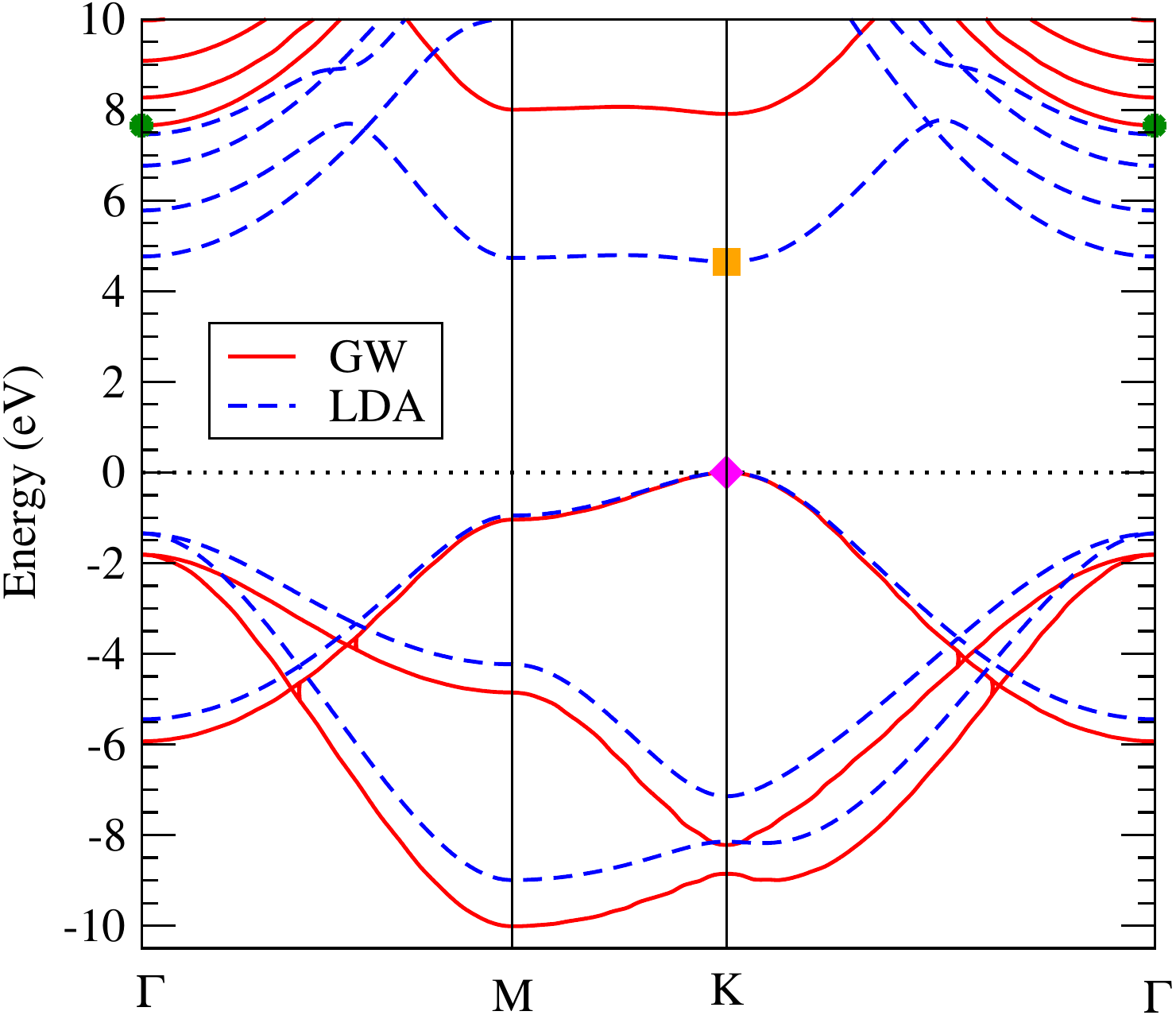}
\caption{Quasiparticle band structure of the single BN sheet from our GW (solid red lines) and
LDA (blue dashed lines) calculations.
Valence band maximum is set at 0 eV, which is at the $K$ point indicated by the pink diamond.
The conduction band minima of the GW and LDA band structures are denoted
by the green star and orange square, respectively.}
\label{fig:band}
\end{figure}

\section{RESULTS AND DISCUSSION}
\subsection{Electronic and optical properties of the single BN sheet}
Both the GW and LDA quasiparticle band structures of the single BN sheet are displayed in Fig. \ref{fig:band}.
Our GW quasiparticle band structure agrees well with the previous 
calculations \cite{Zhang2022,Ahin2009,Rasmussen2016,Galvani2016,Guilhon2019,Mishra2019,Ferreira2019}.
The single BN sheet shares the similar structure with graphene, the honeycomb lattice. However, unlike graphene 
which is a semimetal with $ \pi $ and $ \pi^* $ bands degenerate at the $K$ point,
here the $ \pi $ and $ \pi^* $ bands are well separated due to the ionicity difference 
between B and N \cite{Ismail-Beigi2006}. 
Hence, the single BN sheet has a large band gap (Fig. 2). 
Table I shows that the single BN sheet is an insulator with a $ K^v \rightarrow K^c $ direct band gap 
of 4.65 eV from the LDA calculation. 
However, the single BN sheet becomes an insulator with a larger $ K^v \rightarrow \Gamma^c $ indirect band gap 
of 7.66 eV from the GW calculation. The large GW quasiparticle correction to the LDA band gap can be attributed to the 
weak dielectric screening, a consequence of 2D nature and the wide band gap of the BN sheet. 
The GW quasiparticle correction of the  $ K^v \rightarrow K^c $ gap (3.26 eV) is slightly (0.37 eV) larger than 
that of $ K^v \rightarrow \Gamma^c $ (2.89 eV). This makes the single BN sheet transform 
from the direct band gap ($ K^v \rightarrow K^c $) in the LDA  to the indirect bandgap 
($ K^v \rightarrow \Gamma^c $) at the GW level. The $ \pi^* $ band, arising from $ 2p_z $ orbitals, 
has the out-of-plane charge density  and hence the weaker dielectric screening. The weaker dielectric screening 
of the $ \pi^* $ band explains why the quasiparticle correction of $ K^v \rightarrow K^c $ gap 
is smaller than that of $ K^v \rightarrow \Gamma^c $ gap \cite{Hsueh2011}. 
Consequently, the GW quasiparticle correction not only modifies the band gap, but also changes the dispersion of the bands. 
This indicates the significance of GW quasiparticle calculations, and thus the complex screening effect 
cannot be fully taken into account through a simple scissor correction. 
We notice that our GW band widths are in nearly perfect agreement with the experimental values (Table I).
\begin{table*}[t]
\caption{Transition energies from the valence band maximum ($v$) to conduction band minimum ($c$) ($\Delta E$), 
band gaps ($E_g$), band widths, the onset of the continuum optical absorption ($E_{abs}$), 
exciton excitation energy (i.e., optical band gap) ($\Omega$) and exciton binding energy ($E_b$) of single BN sheet.
The available experimental data (Exp.) are also listed for comparison.}
\begin{ruledtabular}
\begin{tabular}{c c c c c c c c c}
 & \multicolumn{2}{c}{$\Delta E$ ($E_g$)} (eV) & $E_{abs}$ (eV) & \multicolumn{3}{c}{Band width (eV)} & $\Omega$ (eV) & $E_b$ (eV) \\
 \cline{2-3}\cline{4-4}\cline{5-7}\cline{8-9}
 & $ K^v\rightarrow K^c $ & $ K^v\rightarrow \Gamma^c $ & & $ \pi $ Band & $ \sigma_1 $  Band & $ \sigma_2 $  Band & \multicolumn{2}{c}{Exciton A (B)} \\
 \hline
 LDA    & 4.65 (4.65) & 4.77 & 4.52 & 5.45 & 5.80 & 7.65 & ... & ...  \\
 GW     & 7.91 & 7.66 (7.66) & 7.72 &5.93 & 6.41 & 8.20  & ...  & ... \\
 GW+BSE & ...  & ...  & ...  & ...  & ...  & ...  & 5.91 (6.76) & 1.81 (0.96) \\
 Exp.   & ...  & ...  & ...  & 5.8$^a$ & 6.5$^a$ & 8.2$^a$ & 6.05$^b$, 6.03$^c$, 6.1$^d$ & ... \\
\end{tabular}
\end{ruledtabular}
{$^a$Reference \cite{Nagashima1995}; $^b$Reference \cite{Elias2019}; $^c$Reference \cite{Li2017}; $^d$Reference \cite{Roman2021}}
\label{table:band exp}
\end{table*}
%\twocolumngrid

\begin{figure}[tbph] \centering
\includegraphics[width=7.5cm]{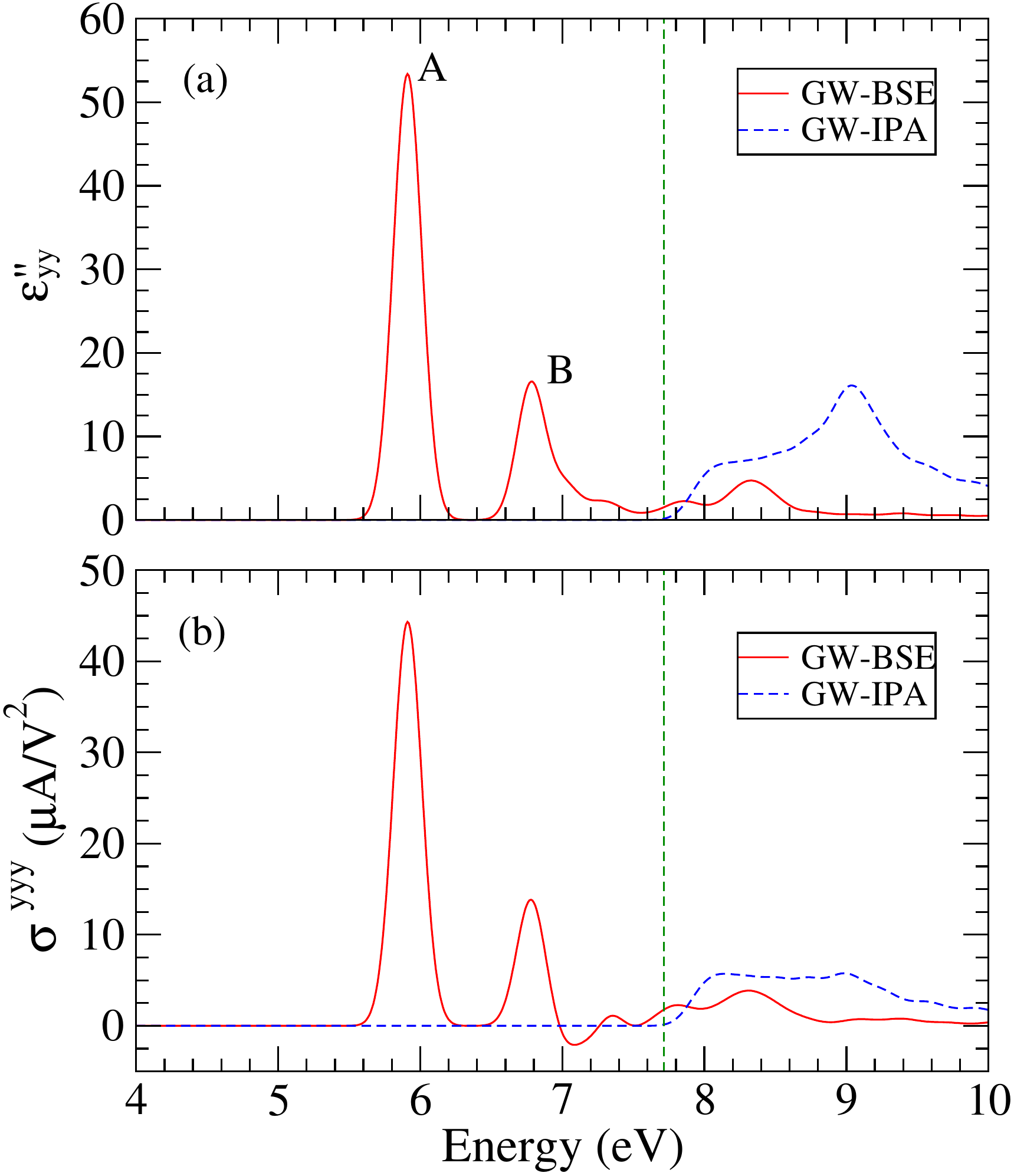}
\caption{(a) Imaginary part of the dielectric function and (b) shift current conductivity 
of the single BN sheet from both GW+BSE (solid red line) and GW+IPA (dashed blue line) calculations. 
The first two bright exciton peaks are labelled by A and B. 
The green vertical line indicates the onset of the continuum optical absorption (see Table I).}
\label{fig:sheet optics}
\end{figure}

Figure \ref{fig:sheet optics}(a) shows the calculated imaginary (absorptive) part ($\varepsilon"$) 
of the dielectric function of the single BN sheet. 
We notice that the $\varepsilon"$ spectrum looks very similar to that of the previous LDA
calculation~\cite{Guo2005a}, {\it albeit} with the onset of the absorption and peak positions
at much higher photon energies due to the GW quasiparticle corrections. 
For optical excitations below 10 eV with the in-plane polarization of the electric field, 
only transitions between the $ \pi $ and $ \pi^* $ bands are dipole allowed. 
The oscillator strength of the dielectric function 
is mainly contributed by the $ \pi $ and $ \pi^* $ bands. The onset energy of the absorption spectrum 
at the GW-IPA level corresponds to the direct transition at the $K$ point. On the other hand, the optical transition 
near the $ M $ point contributes to the largest peak at $\sim$9.0 eV. In other words, the oscillator strength of the dielectric 
function stems primarily from the optical transitions between two flat $ \pi $ and $ \pi^* $ bands along 
the high symmetry $ K-M $ line. 

Remarkably, when the excitonic effect is included at the GW+BSE level,
two prominent absorption peaks [labelled A and B in Fig. 3(a)] occur in the quasiparticle band gap
and they are due to the excitation of the first two bright excitions. Unlike bulk semiconductors
such as CdS in which the exciton peak usually appears a shoulder on the absorption edge~\cite{Sotome2021},
the A and B exciton peaks in the single BN sheet are completely detached from the quasiparticle absorption edge
due to their gigantic binding energies of 1.81 and 0.96 eV (Table I), respectively. 
Note that the calculated large excitation energy of the A exciton 
agrees well with the measured optical band gap (see Table I).

\begin{figure}[tbph] \centering
\includegraphics[width=7.5cm]{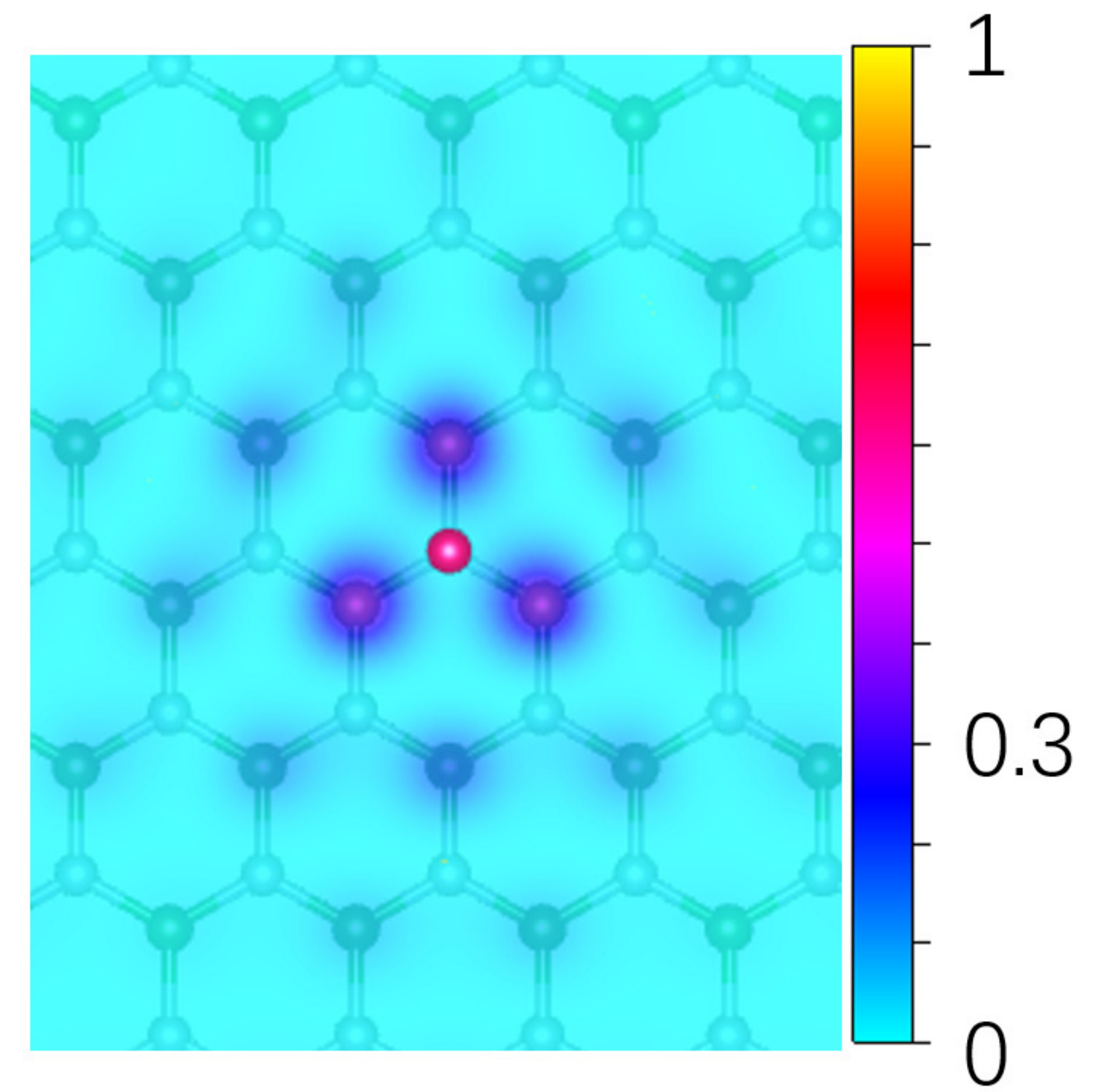}
\caption{The electron probability distribution $ |\psi(\textbf{r}_e,\textbf{r}_h)|^2 $ [see Eq. (\ref{eq2})]
of the doubly degenerate A exciton in the single BN sheet with the hole position ($\textbf{r}_h$)
fixed at the N atom (the pink sphere).}
\label{fig:sheet exciton}
\end{figure}

The large exciton binding energies are the consequence of the weak screening in 2D 
wide bandgap materials such as the single BN sheet, as mentioned in the previous paragraph. 
Figure \ref{fig:sheet exciton} shows the exciton A wavefunction 
($ |\psi(\textbf{r}_e,\textbf{r}_h)|^2$).                            
Clearly, the exciton is strongly localized, and the electron probability distribution
beyond the nearest neighbor is lower than $ 30\% $. 
Consequently, the exciton envelope function of the A exciton 
is widespread in the BZ~\cite{Zhang2022}, resulting in the mixing between 
the interband electron-hole excitations along the high-symmetry $ K-M $ line. The mixing strongly 
redistributes the oscillator strength of the spectrum [see Eq. (\ref{eq4})]. The in-gap exciton states A 
and B carry almost all the oscillator strength, and the absorption spectrum almost diminishes above the 
quasiparticle band gap. 
The magnitude of the A exciton peak is more than three times larger than the largest quasiparticle absorption
peak at $\sim$9.0 eV at the GW-IPA level. 

Figure \ref{fig:sheet optics}(b) displays the calculated shift current conductivity of the BN sheet. 
Clearly, the shift current conductivity spectra are rather similar to the corresponding absorption spectra.
In particular, when the excitonic effect is included,
two huge peaks occur within the quasiparticle band gap (red line). Comparing the shift current conductivity to
the imaginary part of the dielectric function indicates that the two peaks are due to
the A and B exciton excitations. The height of the maximal peak at 5.91 eV is nearly eight times larger
than that of the shoulder at 8.15 eV from the GW-IPA calculation,
i.e., the effect of the exciton excitation not only create two in-gap peaks
but also greatly enhance the shift current. 
The significant increase in the shift current 
is primarily due to the in-gap A exciton excitation, and 
is mainly caused by the significant enhancement of the optical dipole matrix element [see Eq. (\ref{eq4}) and (\ref{eq9})]
due to strong overlap of wavefunctions of the electron and hole within the electron-hole pair.

\subsection{Electronic properties of BN nanotubes}

\begin{figure}[tbph] \centering
\includegraphics[width=7cm]{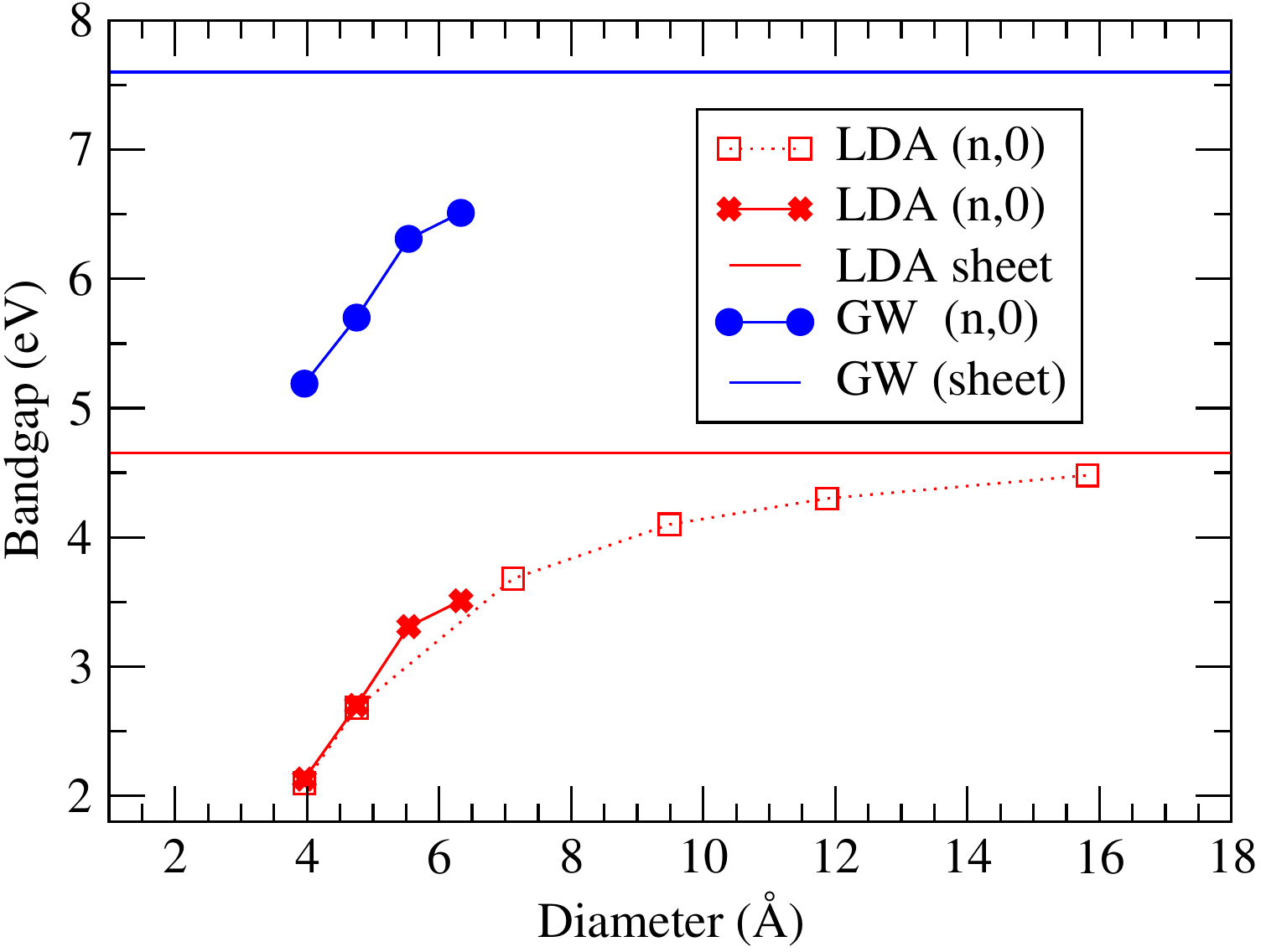}
\caption{Quasiparticle bandgaps of the zigzag BN-NTs vs the tube diameter 
from both LDA and GW calculations. Bandgaps from the previous LDA calculation~\cite{Guo2005a} 
are denoted by red open squares.
The horizontal red and blue lines indicate the quasiparticle band gaps of the single BN sheet
from the LDA and GW calculations, respectively.
}
\label{fig:BN-NTs bandgap}
\end{figure}

\begin{figure}[tbph] \centering
\includegraphics[width=8.5cm]{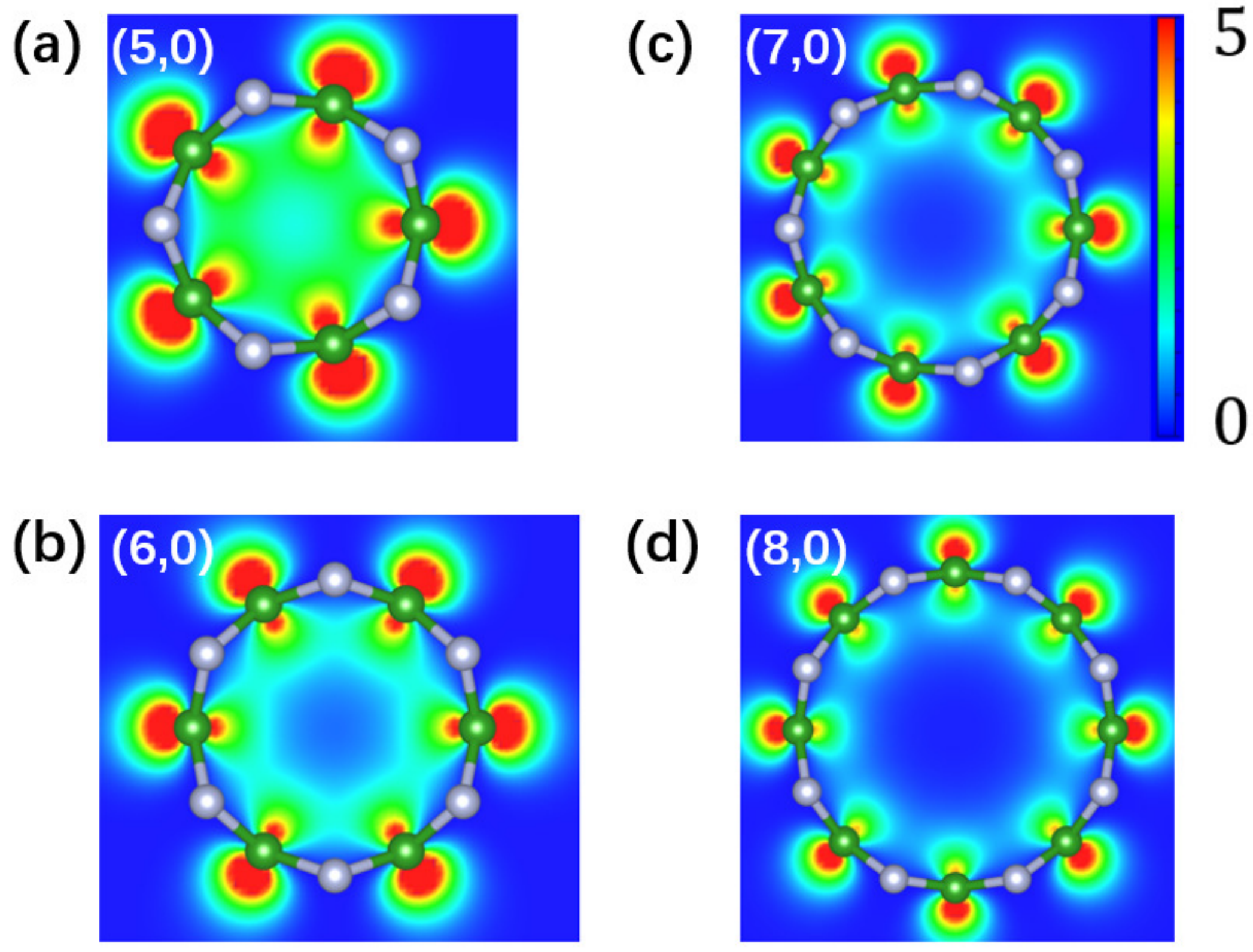}
\caption{Charge density distribution (in units of $10^{-3}$ \AA$^{-3}$) of the states 
in the vicinity of the conduction band minimum of the zigzag [($5,0),(6,0),(7,0),(8,0)$] BN-NTs. 
B and N atoms are indicated by green and gray spheres, respectively.}
\label{fig:CBM charge density}
\end{figure}

Our LDA and GW calculations indicate that the zigzag BN-NTs considered here are direct bandgap insulators 
with $ \Gamma^v \rightarrow \Gamma^c $ transition, being consistent with previous LDA calculations
(see, e.g., Ref. \cite{Guo2005a}). The calculated band gaps of the zigzag BN-NTs are listed 
in Table \ref{table:BN-NT gaps} and also displayed in Fig. \ref{fig:BN-NTs bandgap}. 
Figure \ref{fig:BN-NTs bandgap} shows that for small diameter BN-NTs, both GW and LDA band gaps get reduced
dramatically as the diameter decreases. Interestingly, 
the GW correction to the LDA band gap ($\Delta E_g$) is almost independent of the diameter,
being about 3.0 eV (Table \ref{table:BN-NT gaps} and Fig. \ref{fig:BN-NTs bandgap}).
The lowering of the band gap relative to the BN sheet 
already occurs at the LDA level~\cite{Guo2005a}, and was attributed to 
the curvature-induced orbital rehybridization. The $ \pi^* $ and $ \sigma^* $ orbitals are orthogonal 
to each other in the flat BN sheet. However, such orthogonality does not hold in the BN-NTs 
where the local curvature exists. Consequently, the $ \pi^* $ and $ \sigma^* $ orbitals can hybridize 
and form ring-like charge distribution, as shown in Fig. \ref{fig:CBM charge density}. 
The ring-like charge distribution can effectively reduce the ionicity and hence lowers the energies 
of $ \pi^* $ bands \cite{Blase1994,Reich2004,Rubio1994}, thereby reducing the quasiparticle 
bandgaps of the BN-NTs. As the diameter increases, the effect of curvature-induced orbital rehybridization 
decreases and the band gap converges to that of single BN sheet.

\subsection{Absorption spectra of BN nanotubes}

\begin{figure}[tbph] \centering
\includegraphics[width=8.5cm]{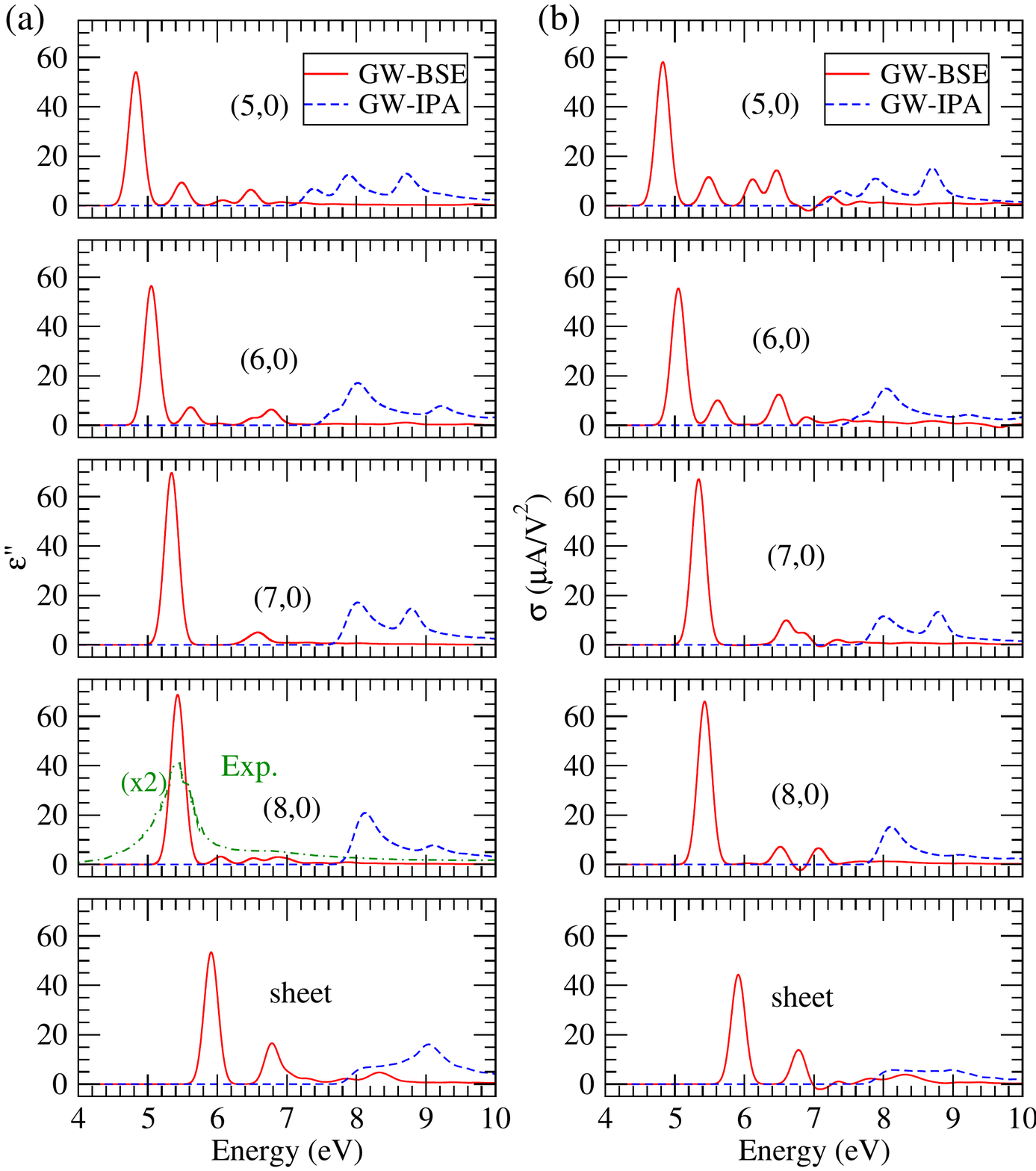}
\caption{(a) The imaginary (absorptive) part of the dielectric function and (b) shift current conductivity 
of the single-walled zigzag [$(5,0),(6,0),(7,0),(8,0)$] BN-NTs and the single BN sheet from
both the GW+BSE (solid red lines) and GW+IPA (dashed blue lines) calculations. The prominent peak
labelled A, is due to the first bright exciton (exciton A). 
Both the polarization direction of light electric field [(a) and (b)] and shift current (b) are parallel 
to the tube axis ($y$-axis) in the BN-NTs (BN sheet).
For comparison, the experimental dielectric function 
of the multi-walled BN-NT~\cite{Fuentes2003} is also displayed in (a).}
\label{fig:BN-NT optics}
\end{figure}

\begin{table*}[t]
\caption{Diameter ($D$), band gap ($E_g$), the onset of the continuum optical absorption ($E_{abs}$),
exciton excitation energy ($\Omega$), exciton binding energy ($E_b$) and the difference 
between the LDA and GW band gaps ($\Delta E^{GW-LDA}_g$) of the zigzag BN-NTs and also single BN sheet (Table I).
For comparison, the measured optical band gaps ($\Omega$) of large diameter BN-NTs are also listed.
}
\begin{ruledtabular}
\begin{tabular}{c c c c c c c c}
 & & & & &\multicolumn{3}{c}{Exciton A}\\
 \cline{6-8}
 & $D$ (\AA) & $E_g^{LDA}$ (eV)& $E_g^{GW}$ (eV) & $\Delta E^{GW-LDA}_g$ (eV) &  $E_{abs}$ (eV) & $E_b$ (eV) & $\Omega$ (eV) \\
 \hline
 BN-NT\\
 (5,0) & 4.11 & 2.13 & 5.19 & 3.06 & 7.06 & 2.23 & 4.83\\
 (6,0) & 4.87 & 2.70 & 5.70 & 3.00 & 7.37 & 2.32 & 5.05\\
 (7,0) & 5.65 & 3.31 & 6.31 & 3.00 & 7.61 & 2.27 & 5.34\\
 (8,0) & 6.43 & 3.51 & 6.51 & 3.00 & 7.71 & 2.28 & 5.43\\
 Exp.$^a$ & $ <10 $ & ... & ... & ... & ... & ... & 5.51 \\
 Exp.$^b$ & 15-30 & ... & ... & ... & ... & ... & 5.8 $ \pm $ 0.2 \\
%Exp.$^c$ & 25 $\pm$ 10 & ... & ... & ... & ... & ... & 5.80 $ \pm $ 0.01 \\
 Exp.$^c$ & 50 $\pm$ 10 & ... & ... & ... & ... & ... & 5.82 $ \pm $ 0.01 \\
 Exp.$^c$ & 600 $\pm$ 100 & ... & ... & ... & ... & ... & 5.90 $ \pm $ 0.01 \\
%Exp.$^c$ & 1000 $\pm$ 200 & ... & ... & ... & ... & ... & 5.90 $ \pm $ 0.01 \\
 sheet & $ \infty $ & 4.65 & 7.66 & 3.01 & 7.72 & 1.81 & 5.91\\
\end{tabular}
\end{ruledtabular}
{$^a$Reference \cite{Li2010}; $^b$Reference \cite{Arenal2005}; $^c$Reference \cite{Yu2009}}
\label{table:BN-NT gaps}
\end{table*}
%\twocolumngrid

We show the calculated absorptive part ($\varepsilon"$)
of the dielectric function of the considered BN-NTs and also single BN sheet in Fig. \ref{fig:BN-NT optics}(a).
In the BN-NTs, optical transitions between $ \pi $ and $ \pi^* $ bands dominate the absorption spectra.
First, compared with the single BN sheet, the energies of the $ \pi^* $ bands of the BN-NTs are lowered owing 
to the curvature-induced orbital rehybridization, as discussed above. Consequently, the energies 
of the $ \varepsilon^{"} $ continuum onset at the GW-IPA level decrease with decreasing diameter, 
as shown in Fig. \ref{fig:BN-NT optics} (a) and Table~\ref{table:BN-NT gaps}. 
Second, contrary to the moderate diameter ($D$ \textgreater 10 \AA) BN-NTs where the $\varepsilon"$
spectrum consists of a single distinct peak \cite{Guo2005a},  the absorption spectra of the small diameter 
BN-NTs considered here feature multiple peaks at the GW-IPA level. For example, the $ \varepsilon^{"} $ 
spectrum of BN-NT (5,0) possesses three distinct peaks at 7.4, 7.9, and 8.7 eV, respectively. 

BN-NTs are 1D wide bandgap insulators. Consequently, due to much
reduced dielectric screening in these 1D wide bandgap materials, 
large exciton binding energies arise. For small diameter BN-NT (5,0), in particular, 
the exciton binding energy $E_b$ of peak A is 2.23 eV, which is larger than that 
(1.81 eV) of the 2D single BN sheet (Table I) and also that (0.63 eV) of the typical 2D
transition metal dichalcogenide semiconductor $\text{MoS}_2$ \cite{Qiu2016}. 
It is also much larger than that of 3D bulk $h$-BN (0.72 eV)~\cite{Arnaud2006} and 2H-MoS$_2$ (0.05)~\cite{Beal1972}. 
Note that the exciton binding energy of typical bulk semiconductor CdS is as small as
0.03 eV~\cite{Sotome2021}. 
The large $ E_b $ results in the emergence of the huge in-gap absorption peak which is
completely detached from the onset of the continuum absorption $\varepsilon^{"}$ spectrum 
of the BN-NT. Most of the oscillator strength of $ \varepsilon^{"} $ 
spectra is carried by the A exciton state of the BN-NTs. 

Table \ref{table:BN-NT gaps} shows the excitation energies ($\Omega$) of exciton A in BN-NTs. 
Our calculated $ \Omega$ converges to that of the single BN sheet 
and agrees well with the experimental values \cite{Arenal2005,Yu2009}. 
We also compare our calculated $ \varepsilon^{"} $ with the experimental one 
derived from electron energy-loss spectra. Experimental $ \varepsilon^{"} $ 
of medium diameter multi-walled BN-NTs with small momentum transfer $ q=0.1$ \AA$^{-1}$ 
is plotted in Fig. \ref{fig:BN-NT optics} (a) (orange dotted line) \cite{Fuentes2003}. 
The complex main peak at 5.5 eV in the experimental $ \varepsilon^{"} $ spectrum may consist 
of exciton A peaks from different walls of the multi-walled BN-NT. 
The substantial broadening of the complex primary peak could be due to 
the consequence of the finite exciton lifetime. The shoulder from around 6.0 to 8.0 eV 
may contribute by minor peaks other than exciton A peak. 

In the low-dimensional BN systems, exciton A is strongly localized, as shown in Fig. \ref{fig:sheet exciton}.
The localized nature of exciton A 
suppresses the curvature effect on the exciton binding energy as the tube diameter increases, 
and leads to the fast convergence of the exciton binding energy [see Table \ref{table:BN-NT gaps}] \cite{Wirtz2006}. 
The red shift of the A peak can be attributed to the curvature-induced orbital 
rehybridization that lowers the energies of $ \pi^* $ bands and reduces the onset 
of the continuum $ \varepsilon^{"} $ at the GW-IPA level. 

\subsection{Exciton shift current of BN nanotubes}
In Fig. \ref{fig:BN-NT optics} (b), calculated shift current conductivity spectra of the BN-NTs 
are plotted. Figure \ref{fig:BN-NT optics} shows that the shift current spectra
are very similar to the corresponding $ \varepsilon^{"} $ spectra. This correlation is facilitated by
the dipole matrix elements present in both the expression for $ \varepsilon^{"} $ [Eq. (\ref{eq4})] 
and the expression for the shift current [Eq. (9)]. 
In particular, as for the $ \varepsilon^{"} $ spectrum, the gigantic excitonic effect results in 
the occurrence of the huge in-gap shift current peak from the GW-BSE calculation 
[Fig. \ref{fig:BN-NT optics} (b)]. The prominent shift current peak in the BN-NTs and 
single BN sheet is due to the excitation of the A exciton.
Furthermore, the excitonic effect substantially increases the largest shift current peak 
of all the BN-NTs by nearly a factor of three. This large enhancement of the shift current 
by the A exciton excitation is mainly due to the large enhancement of the dipole matrix element 
caused by the strong overlap of the wavefunctions of the electron and hole in the A exciton state.

The optical responses of BN-NTs in the low-energy region is dominated by $ \pi $-$ \pi^* $ transitions.
As mentioned before, the curvature-induced orbital rehybridization lowers the energy
of the $ \pi^* $ bands of BN-NTs. As a result, the onset of the shift current spectra at the GW-IPA level 
decreases with decreasing diameter, as shown in Fig. \ref{fig:BN-NT optics} (b).
Similar to $ \varepsilon^{"} $ spectra, shift current spectra of small diameter BN-NTs feature
multiple peaks at the GW-IPA level. 

\begin{figure}[tbph] \centering
\includegraphics[width=8.5cm]{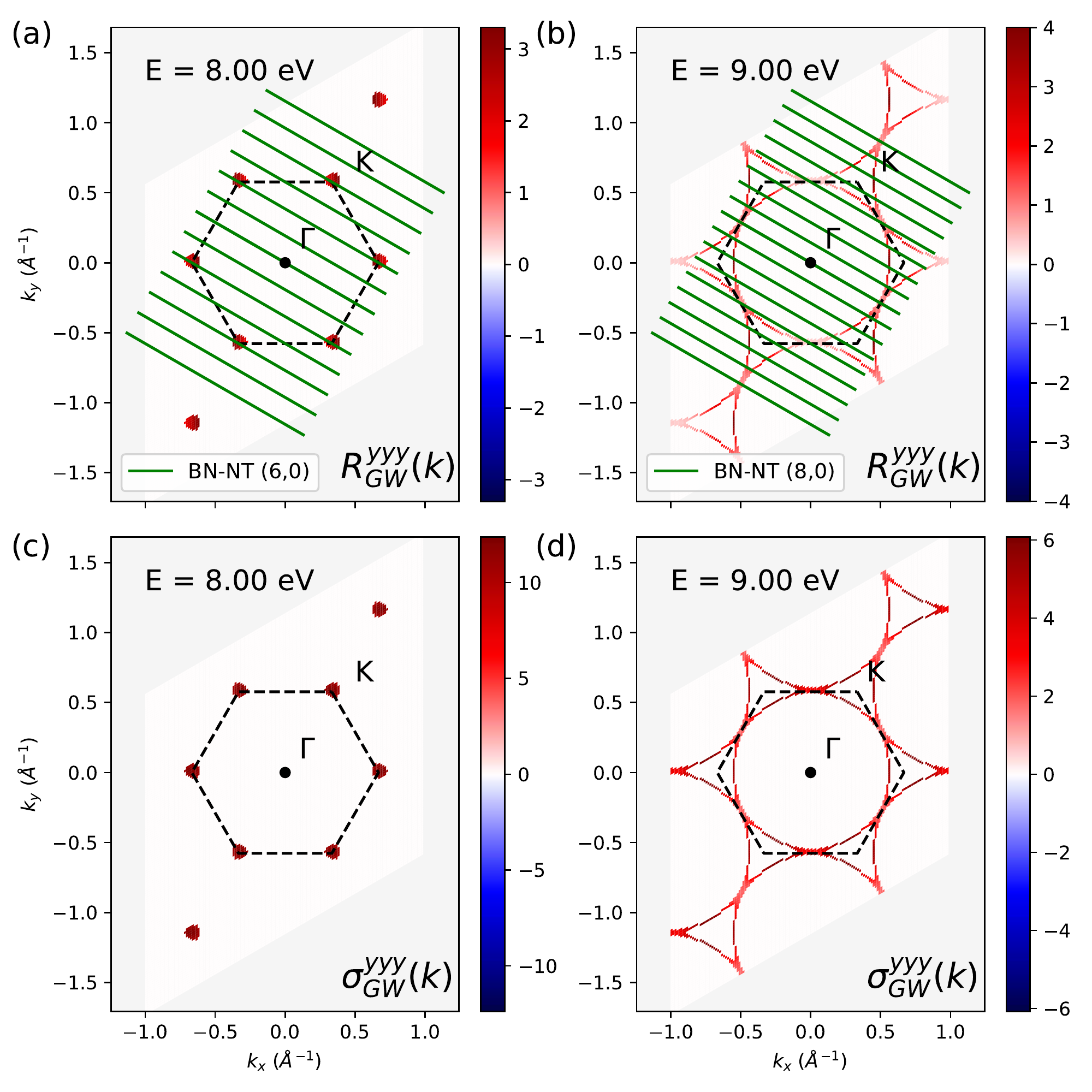}
\caption{
The {\bf k}-resolved shift vector ($R$) (a, b) amd shift current conductivity ($\sigma$) (c, d)
of the single BN sheet from the GW-IPA calculation. The green lines in (a) and (b)
indicate the 1D BZ of the ($6,0$) and ($8,0$) BN-NTs, respectively.
}
\label{fig:k shift current}
\end{figure}

Figure \ref{fig:BN-NT optics} (b) shows that the shift current spectra of the ($n,0$) BN-NTs
from both {\it ab initio} GW-IPA and GW-BSE calculations do not change sign as the tube diameter 
(or $n$) increases, i.e., the direction of the calculated shift current does not follow 
the simple rule of $ sgn(J_\text{shift})=\text{mod}(n,3)$ predicted by the previous tight-binding 
model calculations \cite{Konabe2021,Kral2000}. To better understand this important finding,  
we compute the {\bf k}-resolved shift conductivity and shift vector, which is given
by $ R^{yyy}_{\text{GW-IPA}}(\textbf{k})={\text{Im}}[\textbf{x}^y_{mn}[\textbf{x}^{y}_{nm}]_{;k_y}]/\left|\textbf{x}^y_{mn}\right|^2 $ \cite{Strasser2022},
of the single BN sheet, as displayed in Fig. \ref{fig:k shift current} for two incident photon
energies of 8.0 eV (optical absorption edge) and 9.0 eV (the peak position) at the GW-IPA level. 
Figure \ref{fig:k shift current} indicates that the {\bf k}-resolved shift current and shift vector 
of the single BN sheet are positive in the whole BZ for both selected photon energies. 
At the absorption edge (8.0 eV), the shift current and vector are localized on the $K$ and $K'$ points. 
As the energy increases to the peak position (9.0 eV), the weights of the shift current and shift
vector extend toward the $M$ point and eventually toward the $K$-$\Gamma$ symmetry line.
The 1D BZ of one BN-NT consists of a few discrete lines on the 2D BZ of the BN sheet~\cite{Saito1998},
as displayed for the ($6,0$) and ($8,0$) BN-NTs in Fig. \ref{fig:k shift current}.
Consequently, the direction of the shift current of the zigzag ($n,0$) BN-NTs
would remain unchanged when the tube diameter [or the chirality ($n,0$)] varies,
contrary to the prediction of the tight-binding model calculations \cite{Konabe2021,Kral2000}.  

We notice that previous tight-binding calculations also predicted that the direction of 
the electric polarization of the zigzag BN-NTs would follow the same rule 
of $sgn(J_\text{shift})=\text{mod}(n,3)$ as the shift current \cite{Kral2000,Mele2002}. 
This is not surprising since the electric polarization and shift current are closely connected~\cite{Fregoso2017}. 
However, a subsequent $ab$ $initio$ calculation
showed that the electric polarization of the zigzag BN-NTs grows monotonically as the tube 
diameter increases~\cite{Nakhmanson2003}, i.e., the direction of the electric polarization
does not depend on the chiral index ($n$).
Therefore, both the previous~\cite{Nakhmanson2003} and present {\it ab initio} calculations 
demonstrate the importance of the full {\it ab initio} calculations for the electric 
polarization and shift current in the BN-NTs.

Furthermore, a close examination of the simple tight-binding  model Hamiltonian suggests that
the erroneous prediction of the tight-binding model calculations \cite{Kral2000,Konabe2021}
would stem from the fact that only $\pi$-bands in the single BN sheet and hence the transitions
between the lowest two azimuthal subbands in zigzag BN-NTs were included, whereas the predominant exciton peak
in BN-NTs is composed of a coherent supposition of transitions from several different
subband pairs \cite{Park2006}. Consequently, the rehybridization effect,
induced by the curvature of small diameter BN-NTs, is missing in the simple tight binding
method \cite{Kral2000,Konabe2021}. Indeed, previous 
DFT calculations showed that the rehybridization effect plays a crucial role
in the strong bandgap renormalization for small diameter zigzag BN-NTs \cite{Guo2005a}.

Let us now compare the magnitude of the excitonic photocurrents in the considered BN systems
with the observed shift currents in well-known materials. First, the observed photocurrent
in ferroelectric BaTiO$_3$ above the absorption edge under light intensity $ I=0.5 $ mW/cm$^2$ 
and sample width $ w = 0.15 $ cm is around 5$\times10^{-13}$ A~\cite{Koch1976}.
We find that under the same conditions, the A exciton shift current of a single BN sheet 
would reach $5.0\times10^{-12} $ A, and that from a thin film consisting of, e.g., 
the ($5,0$) BN-NT array with an intertubular distance ($d$) of 3.28 Å would be as large as $\sim$10 mA. 
Second, the effective conductivity of the A exciton shift current in the considered
BN systems (Fig. \ref{fig:BN-NT optics}) is about two-orders of magnitude larger
than the excitonic shift conductivity ($\sim0.2$ $\mu$A/V$^2$) observed in semiconductor CdS~\cite{Sotome2021},
and also more than five times larger than the largest observed shift conductivity
($\sim10.0$ $\mu$A/V$^2$) in ferroelectric SbSI~\cite{Sotome2019}. 
 
\section{CONCLUSIONS}
Using the newly developed computationally efficient algorithms, we have performed 
the state-of-the-art {\it ab initio} GW-BSE culations 
to investigate the exciton shift current as well as  the electronic
and optical properties of the zigzag [$(5,0),(6,0),(7,0),(8,0)$] BN-NTs
and also the single BN sheet. 
First of all, we find a giant in-gap peak in both the shift current and optical absorption spectra
in all the studied BN systems due to the excitation of the A exciton.
This excitonic peak is nearly three times higher than
that in the continuum due to the excitation of the free electron-hole pairs (Fig. 7),
and may be attributed to the gigantic enhancement 
of the optical dipole matrix element by the A exciton resonance. 
Second, our {\it ab initio} calculations show that
the exciton excitation energies and also the onset of the continuum spectra decrease
significantly with the decreasing diameter due to
the curvature-induced orbital rehybridization in small diameter zigzag BN nanotubes.
This orbital rehybridization lowers the ionicity of the $\pi^*$ bands by the formation
of ring-like charge distribution inside the BN-NTs (Fig. 6), thus reducing the energy
of the $ \pi^* $ bands and hence the bandgap.
The quasiparticle GW correction to the bandgap is almost independent of the tube diameter.

Third, we find that the direction of the shift current in the BN-NTs
is independent of the tube chirality ($n,0$) (or diameter), contrary
to the simple detrimental rule of $ sgn(J_\text{shift})=\text{mod}(n,3)$ reported by previous
model Hamiltonian studies \cite{Kral2000,Konabe2021}. Importantly, this implies that 
in a bundle of aligned zigzag BN-NTs, the contributions of the BN-NTs to the shift current 
would be additive rather than cancelling each other as the simple rule suggests~\cite{Kral2000,Konabe2021}. 
Finally, the effective exciton shift current conductivity is nearly ten times larger than 
the largest shift conductivity observed in ferroelectric semiconductors~\cite{Sotome2019}.
Furthermore, strongly bounded excitons with a large binding energy in the considered BN systems
would impede thermal dissociation into free electron-hole pairs. All
these properties would make the zigzag BN-NTs excellent candidates 
for experimentally studying the exciton shift current and also for 
promising applications in nano-scale optoelectronic devices.

\section*{ACKNOWLEDGMENTS}
The authors gratefully acknowledge the support from the National Science and Technology Council 
and the National Center for Theoretical Sciences of The R.O.C. The authors also thank 
the National Center for High-performance Computing (NCHC) in Taiwan for providing computational and storage resources.

\bibliographystyle{apsrev}
\bibliography{TopoCat_abbr}

\end{document}